\documentclass[iop]{emulateapj}
\pdfoutput=1 
\usepackage{hyperref}
\usepackage{amsmath,amstext}
\usepackage[T1]{fontenc}
\usepackage{apjfonts} 
\usepackage[figure,figure*]{hypcap}

\usepackage{color}

\shorttitle{Are LIGO's Black Holes Made From Smaller Black Holes?}
\shortauthors{Fishbach, Holz, \& Farr}

\begin{document}

\title{Are LIGO's Black Holes Made From Smaller Black Holes?}

\author{Maya Fishbach}
\affiliation{Department of Astronomy and Astrophysics, University of Chicago, Chicago, IL 60637, USA}
\author{Daniel E. Holz}
\affiliation{Enrico Fermi Institute, Department of Physics, Department of Astronomy and Astrophysics,\\and Kavli Institute for Cosmological Physics, University of Chicago, Chicago, IL 60637, USA}
\author{Ben Farr}
\affiliation{Enrico Fermi Institute and Kavli Institute for Cosmological Physics,
University of Chicago, Chicago, IL 60637, USA}

\begin{abstract}
One proposed formation channel for stellar mass black holes (BHs) is
through hierarchical mergers of smaller BHs. Repeated mergers
between comparable mass BHs leave an imprint on the spin of the
resulting BH, since the final BH spin is largely
determined by the orbital angular momentum of the binary.
We find that for stellar mass BHs forming hierarchically the distribution of spin magnitudes is
universal, with a peak at $a \sim 0.7$ and little support below
$a \sim 0.5$. We show that the spin distribution is robust against changes to the
mass ratio of the merging binaries, the initial spin distribution
of the first generation of BHs, and the number of merger
generations. While we assume an isotropic distribution of initial spin
directions, spins that are preferentially aligned
or antialigned do not qualitatively change our results. We also consider a ``cluster catastrophe'' model
for BH formation in which we allow for mergers of arbitrary mass
ratios and show that this scenario predicts a unique spin
distribution that is similar to the universal distribution derived for major majors. We explore the ability of spin
measurements from ground-based gravitational-wave (GW) detectors to
constrain hierarchical merger scenarios. We apply a hierarchical
Bayesian mixture model to mock GW data and argue
that the fraction of BHs that formed through hierarchical mergers
will be constrained with $\mathcal{O}(100)$ LIGO binary black hole
detections, while with
$\mathcal{O}(10)$ detections we could falsify a model in which
all component BHs form hierarchically.
\end{abstract}


\section{Introduction}
LIGO's first detections of gravitational waves (GWs) from binary black hole (BBH) systems allow us to probe the formation histories of stellar mass binary black holes (BHs; \citealp{2016ApJ...818L..22A}). Various formation channels have been proposed for the component black holes in these binaries \citep{2016Natur.534..512B, 2016ApJ...824L...8R, 2016MNRAS.460.3545D, 2016PhRvL.116t1301B, CG16, Antonini16, 2017arXiv170104823I}, and these can be broadly separated into two classes: isolated binary evolution and dynamical binary formation channels that involve first-generation BHs (e.g., resulting from stellar collapse) and dynamical formation channels that involve BHs built up from the mergers of earlier generations of BHs. In this work, we consider the latter group: BH formation through hierarchical mergers wherein 
a BH in a BH binary is produced by a merger of two smaller BHs from a previous generation, and the previous generation's BHs may themselves be merger products of an even earlier generation. Hierarchical mergers may occur in high-density environments where some fraction of merger products do not escape despite receiving recoil kicks \citep{Merritt} and may therefore undergo another merger. The hierarchical merger scenario has been proposed in the context of dynamical formation in nuclear star clusters~\citep{Antonini16}, young stellar clusters~\citep{Mapelli16}, AGN disks~\citep{McKernan}, as well as in the formation of primordial BHs ~\citep{CG16}.

In anticipation of future LIGO BBH detections, we describe a method to determine whether or not the observed BHs formed hierarchically. In particular, the hierarchical formation channel can be probed by analyzing the distribution of observed spin magnitudes of the component BHs.

Each BH in a binary has a mass $m_i$ $(i = 1, 2)$ and spin \begin{equation} \mathbf{S_i} = a_i \frac{Gm_i^2}{c}\hat{\mathbf{S_i}}, \end{equation} where $a_i$ is the dimensionless spin magnitude and $\hat{\mathbf{S_i}}$ is the unit spin vector. Because the spins of the BHs in a binary system influence the dynamics of the inspiral and merger, a GW detection provides a measurement of the component spins \citep{2016PhRvX...6d1015A}.  
 
For an individual GW event, the spin measurements are often poorly constrained \citep{2014PhRvL.112y1101V, Purrer}, but we can combine individual spin posteriors to examine the distribution of dimensionless spin magnitudes across all events. In this Letter, we show that the hierarchical merger scenario yields a unique distribution of BH spin magnitudes $a$; therefore, by measuring the spins of observed systems, we can constrain this formation process. Our approach is complementary to that of \cite{Gerosa}, who study the expected distributions of mass, redshift, and binary spin parameter  $\chi_\mathrm{eff}$ for populations of first- and second-generation BHs and show how to use all three measurements to constrain the fraction of second-generation BHs in a detected population. In contrast, we focus solely on GW measurements of spin magnitude $a$ and consider arbitrary generations of previous mergers.

To construct the distribution of BH spin magnitudes resulting from hierarchical mergers, we utilize previous studies of the evolution of BH spins through binary coalescence. Due to advancements in numerical relativity (NR) and post-Newtonian (PN) methods, a number of groups have developed reliable formulae for the final spin following a merger of two spinning BHs~\citep{BKL07, Kesden08,Tichy08,HLZ14,HBR16,Keitel}. Intuitively, there are two contributions to the spin following a coalescence: the individual spins of the two progenitor BHs and the binary system's orbital angular momentum. As the BBHs inspiral toward each other, they lose energy and orbital angular momentum through the emission of GWs. When the Bus finally merge, as shown by \citet{BKL07}, the remaining orbital angular momentum that contributes to the final BH spin can be approximated by the orbital angular momentum of a test particle at the innermost stable circular orbit of the final BH (where the mass of the test particle is taken to be the reduced mass of the BBHs). The contribution from the orbital angular momentum will be most significant for equal mass BBHs and will dominate over the contribution from the spin angular momentum. For example, as is well understood from NR simulations, a merger of nonspinning BBHs of equal mass will result in a final BH with a dimensionless spin magnitude of $0.6864$ \citep{HBR16}. In order for the spins of the BBHs to cancel the orbital angular momentum, resulting in a nonspinning BH, the spins must be sufficiently large and antialigned to the orbital angular momentum, and the mass ratio $q \equiv m_2/m_1 \leq 1$ must be sufficiently small. In fact, using the results of \cite{BKL07}, the antialigned contributions to the spins, $a_1^-$, $a_2^-$, and the mass ratio, $q$, must satisfy
\begin{equation} \frac{1}{q}a_1^- + qa_2^- + 2\sqrt{3} = 0 \end{equation}
in order to end up with a nonspinning BH. Thus, even for maximally antialigned spins, the mass ratio must satisfy $q < \sqrt 3 - \sqrt 2 \approx 0.32$ in order to overwhelm the orbital angular momentum.  As we shall see, this explains why major mergers (in which $q \sim 1$) result in BHs with a relatively high spin distribution, peaked at $a = 0.69$, and with little support below $a \approx 0.5$. \par

In this work, we consider major mergers ($q \gtrsim 0.7$) as the basis of the hierarchical merger scenario. If the BHs of each generation interact with each other dynamically, they are more likely to form binaries with BHs of similar mass \citep{1993Natur.364..423S,2016ApJ...824L...8R} and we would expect mergers of near-equal mass BHs \citep{2016PhRvD..93h4029R, 2016ApJ...824L..12O}. We would similarly expect near-unity mass ratios for BBHs of primordial origin, as PBH formation scenarios generally allow a narrow mass range for the first generation \citep{2016arXiv161101157K}, and we assume that because of dynamical considerations, such BHs only merge with partners of the same generation.
 
The assumption of major mergers differs from the seminal work of \cite{BH03}, which considered the spin evolution of supermassive BHs as they grow through minor mergers. In contrast to major mergers, minor mergers tend to decrease the spin of the final BH, because the binary's orbital angular momentum is smallest when it augments the total BBH spins (a prograde orbit) and largest when it counteracts it (a retrograde orbit). 

We also assume that, in the absence of any aligning mechanism, the spins of each generation of BHs in the hierarchical merger scenario are isotropically distributed on the sphere. The effects of BBH spins that are preferentially aligned or antialigned with the orbital angular momentum are discussed in Section \ref{results}. However, it is important to note that spins that are initially partially aligned (antialigned) with the orbital angular momentum can become significantly antialigned (aligned) during the inspiral due to precession ~\citep{Kesden10}. This will not affect an isotropic distribution of spins, as a distribution of spins that is isotropic at large distances will remain isotropic during the inspiral up to the point of plunge \citep{Kesden10}. Furthermore, the magnitudes of the BBH spins remain nearly constant during the inspiral (up to 2PN order), which further lends confidence to our calculation of the hierarchical merger spin distribution.

\begin{figure}
\label{HMscenario1}
\includegraphics[width=0.5\textwidth]{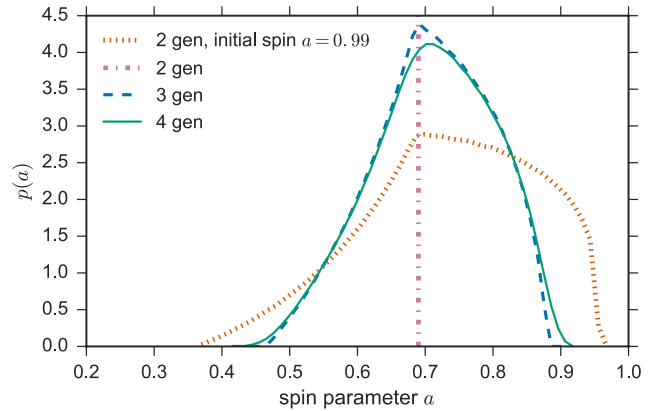}
\capstart
\caption{Probability distribution for the dimensionless spin magnitude for each generation of BHs formed through hierarchical mergers. Unless labeled otherwise, the first generation is nonspinning ($a = 0$) and all mergers take place between equal mass BHs ($q = 1$). For each generation, the spin directions are assumed to be isotropically distributed. Note the rapid convergence to a universal distribution (turquoise solid line). The dotted orange line shows the second-generation distribution for the case where the first generation has near-maximal ($a = 0.99$) spins. The initially nonspinning ($a=0$) and initially near-maximally spinning ($a = 0.99$) cases are indistinguishable by the fourth generation, converging on the universal distribution.}
\end{figure}

\section{Methods}
\subsection{Hierarchical Merger Spin Distribution}
We apply the formulas of \cite{HBR16} to predict the final BH spin from a merger of two BHs, given the spin vectors and masses of the component BHs. This allows us to build a statistical distribution of spin magnitudes resulting from hierarchical mergers, similar to the distributions found by \cite{Tichy08} and \cite{LoustoStatistics}. 

Although we assume major mergers and isotropically distributed spin orientations, we wish to remain general with respect to other aspects of the hierarchical merger scenario. In particular, we do not at the outset specify the spin distribution of the first generation of BHs (before any mergers have occurred) or the exact distribution of mass ratios of merging BHs (although we limit ourselves to $q \geq 0.7$). Furthermore, the desired spin distribution presumably evolves as each generation's BHs merge to form the next generation, but we do not wish to restrict ourselves to a particular generation of the hierarchical merger scenario. Fortunately, as we show below, the resulting spin distribution is relatively insensitive to the spin magnitudes of the first generation, the mass ratios (within the range $0.7 \leq q \leq 1$), or which generation we consider (starting with the second generation). We demonstrate this explicitly by computing spin distributions under various choices of these parameters.

We compute probability density functions of dimensionless spin magnitudes as follows:
we start by taking a large ($6.25 \times 10^6$) ensemble of BHs, and then randomly pick pairs of BHs from this first generation and merge each pair, calculating the final spin from the \cite{HBR16} formula. This gives us the distribution of spin magnitudes for the second generation of BHs. In the simplest case we take the first generation of BHs to be all of the same mass and nonspinning, in which case, the second generation's BHs will all be of roughly double the mass and spinning with dimensionless spin magnitude $a = 0.69$. If the initial generation of BHs is equal mass but with isotropic, near-maximal ($a = 0.99$) spins, the second generation of BHs will have a distribution of spin magnitudes that is similarly peaked at $a \sim 0.7$ with slightly wider support (see Fig. 1). 

To calculate the spin distribution for the third generation of BHs, we randomly and repeatedly choose pairs of BH spin magnitudes from the second generation and randomly choose their spin directions from a spherically isotropic distribution. This yields the spin magnitudes of the third generation of BHs, and we can iterate this procedure to calculate the distribution of BH spins for the $n$th generation given the distribution of spins for the $(n-1)$th generation. In practice, we find that the spin distribution changes only slightly between the third and the fourth generation and has fully converged by the fourth generation, regardless of the initial spin distribution (see Fig.~\ref{HMscenario1}). Different choices of initial spin lead to indistinguishable spin distributions by the third generation. 

\begin{figure}
\label{q_compare}
\includegraphics[width=0.5\textwidth]{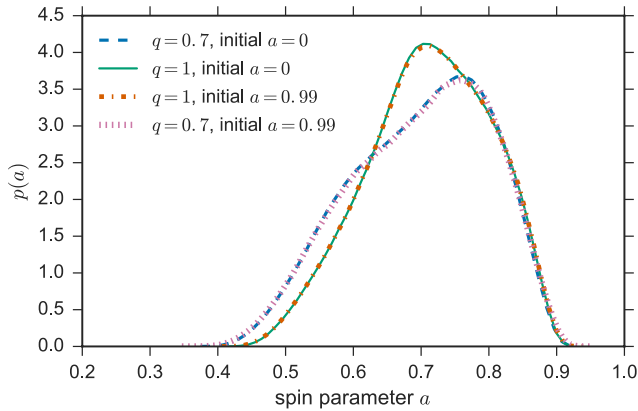}
\capstart
\caption{Converged spin distributions (fourth generation) of hierarchically formed BHs, where in one scenario the BHs always merge with equal mass ($q=1$) and in the other scenario the BHs always merge with mass ratio $q = 0.7$. Changing the spin of the first generation leads to indistinguishable distributions.
}
\end{figure}
To explore the effect of different mass ratios, we consider a toy model in which all mergers occur with a mass ratio of $q=0.7$, instead of $q=1$ as assumed above.
We find that the resulting spin distributions are very similar (see Fig.~\ref{q_compare}), suggesting that any distribution of mass ratios in the range $0.7 \leq q \leq 1$ (as expected for dynamically forming binaries) would not significantly affect the distribution of final spins.
Thus, we find that regardless of mass ratio and initial spin, the hierarchical merger scenario gives rise to a single, standard distribution of BH dimensionless spin magnitudes sharply peaked at $a \sim 0.7$ with nonzero support over $0.4 \lesssim a \lesssim 0.9$. In what follows when we refer to the hierarchical merger spin distribution, we mean the $q = 1$ distribution shown in Fig.~\ref{q_compare}, calculated as the fourth generation of equal mass mergers. An alternate choice would not significantly affect our results.

Our findings are consistent with \cite{Berti}, who found that following a single merger the final spin magnitude is $a \approx 0.7$ regardless of the initial spin magnitude, assuming we average over an isotropic distribution of spin directions (see their Fig.~2). Our results are also consistent with~\cite{Tichy08} and~\cite{LoustoStatistics}, both of whom found that the distribution of spin magnitudes converges after four generations of repeated mergers. However, note that by limiting ourselves to the major mergers relevant for stellar mass BHs, our hierarchical merger spin distribution is different from the distribution presented in Fig.~19 of \cite{LoustoStatistics}, as they considered a wide distribution of mass ratios appropriate for supermassive BHs. In particular, our distribution has little support below $a \sim 0.5$. Our hierarchical merger spin distribution is most similar to the $q =1$ distribution in Fig.~1 of \cite{Tichy08} and Fig.~20 of \cite{LoustoStatistics}; however, we argue that this distribution is insensitive to the initial BH spins and is an adequate description of the second and third generation of BHs even before it fully converges in the fourth generation.

\subsection{Mixture Model Analysis}
We apply a hierarchical Bayesian framework \citep{Hogg2010,Mandel2011} to analyze a collection of BH spin measurements, where the spin measurement from the $i$th GW detection takes the form of a two-dimensional posterior for the BBH spin magnitudes $p(\pmb{\alpha}_i|d_i)$, where $\pmb{\alpha}_i = (a_{1,i}, a_{2,i})$ and $d_i$ is the data. To understand the true population of BH spins from the observed BBHs, we assume that the true spin distribution is parameterized in terms of some parameters, $\mathbf{A}$, which we seek to infer. The true spin distribution, therefore, can be written as $p(\pmb{\alpha}|\mathbf{A})$, and we want to know $p(\mathbf{A}|\mathbf{d})$, where $\mathbf{d} = \{d_i\}$ is the data across all GW detections. We assume that each GW detection is independent, so that \begin{equation} \label{hb1} p(\mathbf{A}|\mathbf{d}) = \prod_i p(\mathbf{A}|d_i). \end{equation} 
Furthermore, we have \begin{equation} \label{hb2}
p(\mathbf{A}|d_i) = \int d\pmb{\alpha}_i p(\mathbf{A},\pmb{\alpha}_i|d_i),
\end{equation}
and applying Bayes's rule gives \begin{equation} \label{hb3}
p(\mathbf{A},\pmb{\alpha}_i|d_i) \propto p(d_i|\pmb{\alpha}_i)p(\pmb{\alpha}_i|\mathbf{A})p(\mathbf{A}),
\end{equation}
where $p(d_i|\pmb{\alpha}_i)$ is the two-dimensional likelihood for the BBH spins, and $p(\mathbf{A})$ is the prior probability for the population parameters $\mathbf{A}$. Before we have learned anything about the population distribution of spin magnitudes $p(\pmb{\alpha}_i|\mathbf{A})$, in the analysis of individual events, we assume a two-dimensional flat prior on $\pmb{\alpha}_i$, so that the likelihood $p(d_i|\pmb{\alpha}_i)$ is proportional to the posterior $p(\pmb{\alpha}_i|d_i)$.
Putting together equations \ref{hb1}--\ref{hb3}, we have \begin{equation} \label{hbfinal}
p(\mathbf{A}|\mathbf{d}) \propto \prod_i \left[ \int d\pmb{\alpha}_i p(d_i|\pmb{\alpha}_i)p(\pmb{\alpha}_i|\mathbf{A})\right]p(\mathbf{A})
\end{equation}
where, because $p(d_i|\pmb{\alpha}_i) \propto p(\pmb{\alpha}_i|d_i)$ for a single event, we can evaluate the above integral over $\pmb{\alpha}_i$ of $p(\pmb{\alpha}_i|\mathbf{A})$ weighed by the likelihood $p(d_i|\pmb{\alpha}_i)$ as an average over $N_i$ posterior samples $\pmb{\alpha}_i^k$: 
\begin{align} \label{posteriorsample}
\begin{split}
\int d\pmb{\alpha}_i p(d_i|\pmb{\alpha}_i)p(\pmb{\alpha}_i|\mathbf{A}) &=
\langle p(\pmb{\alpha}_i|\mathbf{A})\rangle_{\pmb{\alpha}_i} \\
&\approx \frac{1}{N_i}\sum_{k=1}^{N_i} p(\pmb{\alpha}_i^k|\mathbf{A})
.\end{split}
\end{align} 
\par In our case, in order to investigate whether the detected BBHs favor the hierarchical merger scenario, we write the true spin population as a mixture model. Lacking a strong astrophysical prior on the distribution of BH spins ~\citep{MillerMiller}, we take some fraction $f_u$ of the BHs to be uniformly spinning over the allowed range $[0,1]$, and the remaining $(1-f_u)$ of the BHs to come from the hierarchical merger population. 
It is straightforward to consider alternate spin magnitude distributions, and the same analysis would apply if we included an additional component in the mixture model or replaced the uniformly distributed component with a different spin distribution. For the mixture model with parameter $f_u$, we have \begin{equation}
p(a|f_u) = 
\begin{cases}
f_u  + (1-f_u)p_\text{hm}(a) & 0\leq a \leq 1 \\
0 & \text{otherwise}
\end{cases}
\end{equation}
where $p_\text{hm}(a)$ is the hierarchical merger spin distribution.
We assume the spins of the BBHs in a single system are independent of one another, so \begin{equation}
p(\pmb{\alpha}_i|f_u) = p(a_{1,i}|f_u)p(a_{2,i}|f_u).
\end{equation}
We also use a flat prior for the mixture parameter
\begin{equation}
p(f_u) =
\begin{cases}
1 & 0 \leq f_u \leq 1 \\
0 & \text{otherwise}.
\end{cases}
\end{equation}
Then for the mixture model, equation \ref{hbfinal} becomes
\begin{align}
\label{pfu1}
\begin{split}
p(f_u|\mathbf{d}) \propto p(f_u) \prod_i \int d\pmb{\alpha}_i p(d_i|\pmb{\alpha}_i) &\left[ f_u + (1-f_u)p_\text{hm}(a_{1,i}) \right] \\
&\cdot \left[ f_u + (1-f_u)p_\text{hm}(a_{2,i}) \right]
\end{split}
\end{align}
Using equation \ref{posteriorsample}, given $N_i$ posterior samples for each BBH, we can approximate equation \ref{pfu1} as
\begin{align}
\begin{split}
p(f_u|\mathbf{d}) \propto p(f_u) \prod_i f_u^2 &+f_u(1-f_u) \frac{1}{N_i} \sum_{k=1}^{N_i} \left[ p_\text{hm}(a_{1,i}^k)+p_\text{hm}(a_{2,i}^k) \right] \\
&+(1-f_u)^2 \frac{1}{N_i} \sum_{k=1}^{N_i} p_\text{hm}(a_{1,i}^k)p_\text{hm}(a_{2,i}^k).
\end{split}
\end{align}
The mixture model parameterization provides a convenient way to compare the hierarchical merger model to any other model (in our case, a model that yields a flat distribution of spins). As \cite{Vitale15} discuss, for a mixture of two (or more) models, we can write the posterior of the mixture parameter in terms of the Bayesian evidence for the models under consideration. In our case, we can relate the posterior $p(f_u|\mathbf{d})$ to the evidence ratio, or Bayes factor, between the hierarchical merger model $\mathcal{H}_\text{hm}$ and the uniform spin model $\mathcal{H}_u$. The evidence for each model given data $d_i$ is defined as $Z^i_\text{hm} = p(d_i|\mathcal{H}_\text{hm})$ and $Z^i_{u} = p(d_i|\mathcal{H}_{u})$, and, assuming GW detections are independent, we can write \begin{align}
p(f_u|\mathbf{d}) &\propto p(\mathbf{d}|f_u)p(f_u) \\
&= p(f_u)\prod_i p(d_i|f_u) \\ 
&= p(f_u) \prod_i \left[ p(d_i,\mathcal{H}_u|f_u)+p(d_i,\mathcal{H}_\text{hm}|f_u) \right] \\
&= p(f_u) \prod_i \left[ p(d_i|\mathcal{H}_u)p(\mathcal{H}_u|f_u)+p(d_i|\mathcal{H}_\text{hm})p(\mathcal{H}_\text{hm}|f_u) \right]\\
&=p(f_u) \prod_i \left[ 
Z^i_uf_u+Z^i_\text{hm}(1-f_u)
\right].
\end{align}
We therefore have that the Bayes's factor is
\begin{equation}
\prod_i\frac{Z^i_u}{Z^i_\text{hm}} = \frac{p(f_u=0)}{p(f_u=1)}
\frac{p(f_u=1|\mathbf{d})}{p(f_u=0|\mathbf{d})}.
\end{equation}
Thus, computing $p(f_u|\mathbf{d})$ allows us to directly find the Bayes's factor, which allows us to argue (or refute) that a population of observed BHs came from the hierarchical merger formation channel. The mixture model is also useful to constrain the fraction of the observations that are consistent with having formed through hierarchical mergers. In the next section, we demonstrate that this analysis will yield meaningful constraints within just a few years of advanced LIGO operation.

\begin{figure}
\label{pfu_sigma_compare}
\includegraphics[width=0.5\textwidth]{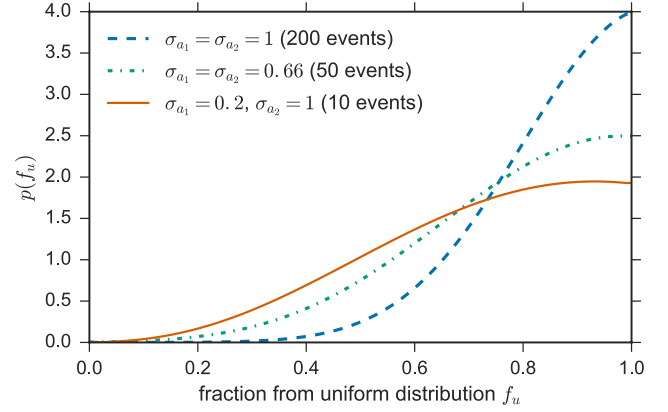}
\capstart
\caption{Posterior probability density functions on the parameter $f_u$ for 3 simulated populations of BHs that have not formed hierarchically (the true $f_u=1$). The populations of BHs all have spin magnitudes drawn from a uniform $[0,1]$ distribution, but differ in the uncertainty on their measured spins magnitudes. We approximate spin measurements as truncated Gaussians and vary $\sigma_{a_1}$, $\sigma_{a_2}$ between populations. We note that 10 detections may be sufficient to rule out a pure hierarchical merger model.}
\end{figure}

\section{Results}
\label{results}
\begin{figure}
\label{pfu_s66}
\includegraphics[width=0.5\textwidth]{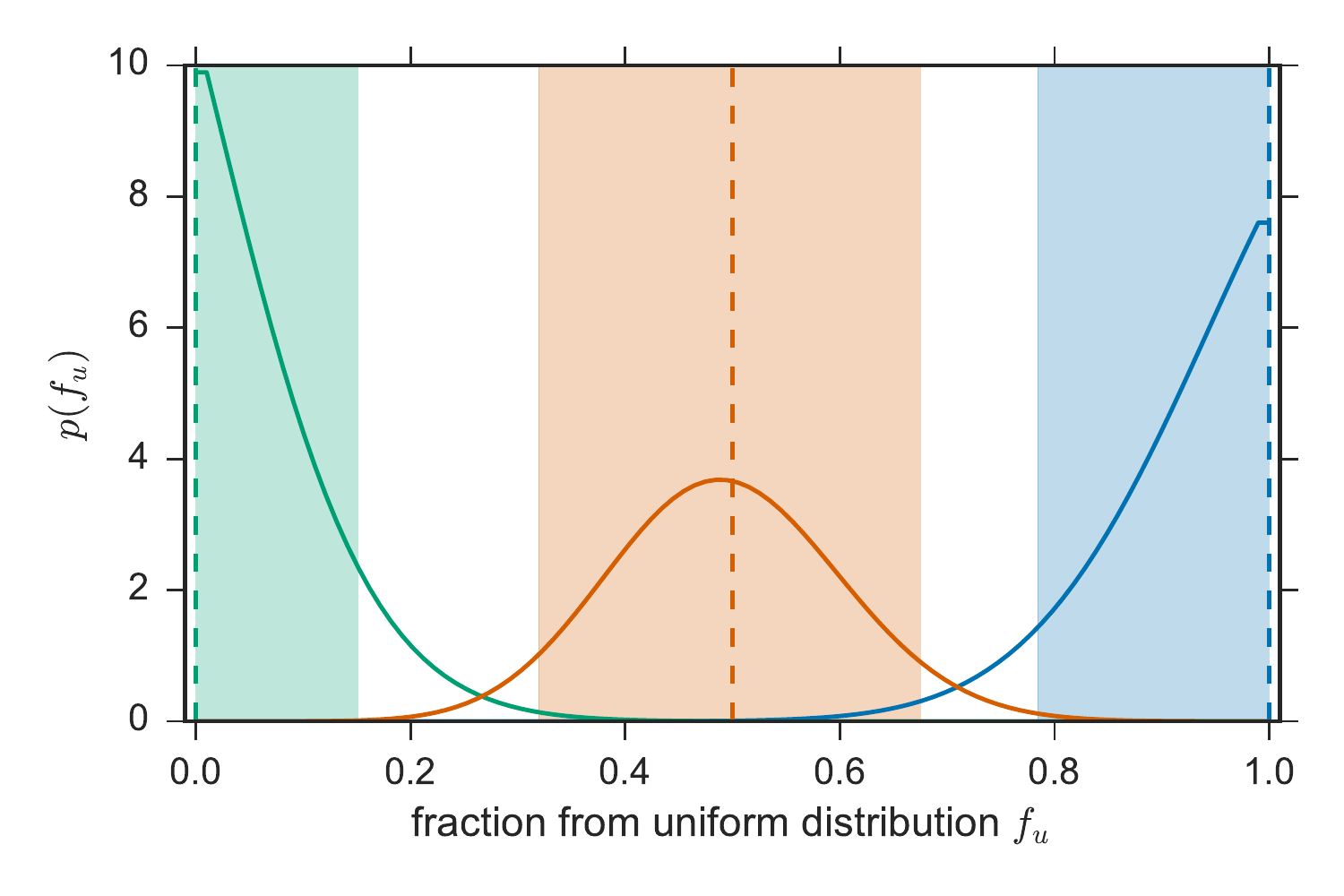}
\capstart
\caption{Posterior probability density functions on the parameter $f_u$ for 3 simulated populations with different values of $f_u$ (given by the dashed lines). The simulated BH populations with $f_u = 0$ (turquoise) and $f_u = 1$ (sky blue) each consist of 200 BBH events, while the simulated BH population with $f_u = 0.5$ (orange) consists of 400 BBH events. We assume that for all events the spin magnitude measurements $p(a_1|d_i)$, $p(a_2|d_i)$ are truncated Gaussians with standard deviations $\sigma = 0.66$. The shaded regions denote $90\%$ credible intervals.}
\end{figure}

For the purposes of this work, we illustrate our method on very simplified spin posterior distributions, leaving the analysis of real data to the LIGO collaboration. We assume that each BBH detection provides a measurement of the two component BH spin magnitudes with some uncertainty \citep[see, for example, Fig.~5 of][]{2016PhRvX...6d1015A}. We neglect correlations between the two spin measurements, which is equivalent to setting \begin{equation}
p(d_i|\pmb{\alpha}_i) = p(d_i|a_{1,i})p(d_i|a_{2,i})
\end{equation}
in equation \ref{pfu1}.
Following \cite{Stevenson17}, we approximate each spin magnitude posterior $p(a_{j,i}|d_i)$ $(j=\{1,2\})$ as a Gaussian, restricted to the range [0,1], with a standard deviation $\sigma_{j,i}$ corresponding to the measurement uncertainty. In other words, for a BH with dimensionless spin magnitude $a_{j,i}^\text{true}$, we generate a posterior centered on $a_{j,i}^\text{data} = a_{j,i}^\text{true} + e_{j,i}$ where $e_{j,i}$ is a random measurement error chosen from $N(0, \sigma_{j,i})$. The spin magnitude posterior for a single BH is then given by
\begin{equation}
\label{fakeposterior}
p(a_{j,i}|d_i) = N(a_{j,i}^\text{data},\sigma_{j,i})
\end{equation} truncated and normalized to our prior range $[0,1]$. With these assumptions, we compute equation \ref{pfu1} by drawing 1000 samples from each spin magnitude posterior given by equation \ref{fakeposterior} for a simulated population of $a_{j,i}^\text{true}$. In other words, we solve
\begin{equation}
p(f_u|\mathbf{d}) \propto p(f_u) \prod_i \prod_{j=1,2} f_u+(1-f_u)\langle p_\text{hm}(a_{j,i}) \rangle_{a_{j,i}}
\end{equation}
where 
\begin{equation}
\langle p_\text{hm}(a_{j,i}) \rangle_{a_{j,i}} \approx \frac{1}{1000}\sum_{k=1}^{1000} p_\text{hm}(a_{j,i}^k)
\end{equation}
for $a_{j,i}^k \sim p(a_{j,i}|d_i)$. 

The uncertainty $\sigma$ on spin magnitude depends on various factors, including the true spin magnitudes, the signal-to-noise ratio (SNR) of the inspiral and ringdown, the mass ratio of the binary, and the orientation of the spin vectors. As demonstrated by \cite{Purrer}, we expect this uncertainty to be rather large and not particularly dependent on the SNR, especially for events where there is little power in the ringdown, so we carry out our analysis with the conservative choice of $\sigma = 1.$ We then repeat the analysis under the assumption that all events are like GW150914 in terms of spin magnitude uncertainty: consisting of one relatively well measured BH spin magnitude with $\sigma = 0.2$, and one poorly measured BH spin magnitude with $\sigma = 1$. This can be expected for events with moderately high SNR in both the inspiral (which constrains the weighted aligned spin combination $\chi_\mathrm{eff}$) and the ringdown (which constrains the spin of the final BH $a_f$). Motivated by the choice of posterior uncertainties in \cite{Stevenson17} and an examination of mass ratio uncertainties and covariances for published LIGO events \citep[see Table~1 and Fig.~4 of][]{2016PhRvX...6d1015A}, we repeat our analysis for spin posteriors with $\sigma = 0.66$.

Our results are similar for all choices of spin magnitude uncertainties, suggesting that LIGO will be able to clearly distinguish between a population of hierarchically formed BHs and a population of uniformly spinning BHs with $\mathcal{O}(100)$ detections (see Fig.~\ref{pfu_s66}), although this will be possible with as few as 10 detections if at least one spin component is relatively well-measured ($\sigma \approx 0.2$) as in the case of GW150914 (see Fig.~\ref{pfu_sigma_compare}). If the true population of detected BHs is mixed, it requires more detections to precisely measure the fraction $1 - f_u$ that have spin magnitudes consistent with formation through hierarchical mergers. We see in Fig.~\ref{pfu_s66} that the $90\%$ confidence interval for $f_u$ is relatively wide for a mixed population with 400 events, although if we are simply interested in ruling out $f_u = 0$ or $f_u = 1$,  $\mathcal{O}(100)$ detections is sufficient even in the most pessimistic case considered. It is straightforward to extrapolate these results to a greater number of detections: as expected, the width of the posterior $p(f_u|\mathbf{d})$ decreases with the number of detections, $N$, as $1/ \sqrt N$.

\begin{figure}[b]
\label{merge-all}
\includegraphics[width=0.5\textwidth]{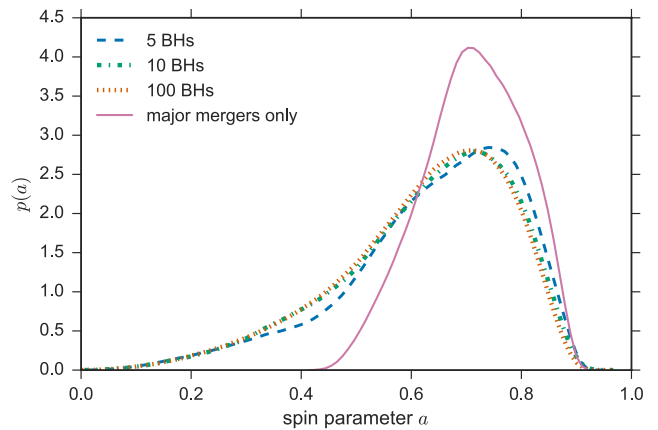}
\capstart
\caption{Probability density function of the spin magnitude of a final BH formed through a ``cluster catastrophe'' of a fixed number of BHs, compared to the universal hierarchical merger distribution. Initially the $N$ BHs are of equal mass and spin magnitudes drawn uniformly on $[0,1]$, and they merge with each other in randomly selected pairs (with isotropic spin directions) until they have all merged into a single final BH.}
\end{figure}
We have thus far analyzed the characteristic spin distribution resulting from major ($q \geq 0.7$) mergers of isotropically spinning BHs. Below, we discuss the implications of relaxing these assumptions. We might expect deviations from isotropic spins in certain astrophysical situations such as BBH formation in a gas-rich AGN disk, where the spins of the component BHs can be preferentially aligned or antialigned with the orbital angular momentum \citep{McKernan}.
If the spins of BBHs are always aligned with the orbital angular momentum, it is straightforward to see that this will result in an even narrower spin distribution, strongly peaked at very high spin magnitudes, converging to $a \sim 0.9$ by the third generation. This situation will thus be easier to constrain with LIGO data.
If both component spins are always antialigned with the orbital angular momentum and the first generation has moderately high spins $a \sim 0.7$, the future generations will have spins $0.43 < a < 0.54$. Even assuming maximal antialigned initial spins, while the second-generation products will be spinning at $0.32 < a < 0.36$ for mass ratios $q \geq 0.7$, the spin distribution will converge to $a \sim 0.5$ starting with the third generation. If the spins are equal in magnitude but one is aligned and the other antialigned with the orbital angular momentum, the situation is identical to a merger of nonspinning BHs and yields a spin magnitude $a \sim 0.7$ (although note that the best-constrained spin parameter $\chi_\mathrm{eff}$ for the binary will be $\chi_\mathrm{eff} = 0$). We conclude that hierarchical major mergers of BBHs cannot produce low spin magnitudes ($a \lesssim 0.4$). 

We can relax the assumption of major mergers by considering an alternative ``cluster catastrophe'' formation scenario in which a fixed number, $N$, of equal mass BHs repeatedly merge in randomly chosen pairs, irrespective of the mass ratio, until there is a single remaining BH. We take the initial distribution of spin magnitudes to be uniform in $[0,1]$ and spin directions to always be isotropic. We find that the spin magnitude of the single remaining BH is insensitive to $N$ or the initial spin magnitudes and is distributed according to the probability distribution in Fig.~\ref{merge-all}. While such a scenario can lead to low mass-ratio mergers in which the primary spin may cancel the orbital angular momentum and produce spin magnitudes $a < 0.4$, these low spin magnitudes remain unlikely: spin magnitudes $a < 0.4$ are produced less than $10\%$ of the time. 

\section{Summary}
We have shown that if BHs build up through hierarchical major mergers of smaller BHs, the spin magnitudes of the resulting BHs follow a universal distribution (see Fig. \ref{q_compare}).
Because this distribution is relatively independent of the  details of the hierarchical merger scenario, we can use it to test whether an observed population of BHs was formed through hierarchical mergers. Although a GW observation of a single coalescing BBH does not strongly constrain individual BH spins \citep{2016PhRvX...6d1015A}, we estimate that the hierarchical formation channel will be strongly constrained with $\mathcal{O}(100)$ detections. Furthermore, we have shown that hierarchical mergers rarely produce BHs with spins below $a \lesssim 0.5$, and even in extreme scenarios that favor antialigned spins or result in a ``cluster catastrophe," one rarely finds BH spins below $a \lesssim 0.4$. If BHs do not form through hierarchical mergers and the spin distribution is uniform on [0,1], we have shown that it will be possible to rule out the hierarchical formation channel with a sample of $\mathcal{O}(10)$ detections (see Fig.~\ref{pfu_sigma_compare}). If instead we select between a hierarchical model and one that favors low spins (instead of uniform as done above), even fewer detections would be sufficient to falsify either model. We note that the spin constraints of the primary component BHs of GW150914 and LVT151012 appear to favor spins $a \lesssim 0.5$ over spins $a \sim 0.7$ (see Figs.~5 of \citealp{PEpaper}, \citealp{2016PhRvX...6d1015A}), suggesting that they are unlikely to have formed through hierarchical major mergers. We leave a quantitative analysis of these events for future work.

\acknowledgments
We thank Davide Gerosa and Emanuele Berti for valuable discussions, and Will Farr and Simon Stevenson for their insight into hierarchical Bayesian mixture models. The authors were supported by NSF CAREER grant
PHY-1151836. They were also supported in part by the Kavli Institute for
Cosmological Physics at the University of Chicago through NSF grant PHY-1125897
and an endowment from the Kavli Foundation.

\bibliographystyle{yahapj}
\bibliography{references}

\begin{thebibliography}{}
\providecommand\natexlab[1]{#1}
\providecommand\JournalTitle[1]{#1}

\bibitem[{{Abbott} {et~al.}(2016{\natexlab{a}}){Abbott}, {Abbott}, {Abbott},
  {Abernathy}, {Acernese}, {Ackley}, {Adams}, {Adams}, {Addesso}, {Adhikari},
  \& et~al.}]{2016ApJ...818L..22A}
{Abbott}, B.~P., {Abbott}, R., {Abbott}, T.~D., {et~al.} 2016{\natexlab{a}},
  \href{http://dx.doi.org/10.3847/2041-8205/818/2/L22}{\JournalTitle{\apjl},
  818, L22}

\bibitem[{{Abbott} {et~al.}(2016{\natexlab{b}}){Abbott}, {Abbott}, {Abbott},
  {Abernathy}, {Acernese}, {Ackley}, {Adams}, {Adams}, {Addesso}, {Adhikari},
  \& et~al.}]{2016PhRvX...6d1015A}
---. 2016{\natexlab{b}},
  \href{http://dx.doi.org/10.1103/PhysRevX.6.041015}{\JournalTitle{Physical
  Review X}, 6, 041015}

\bibitem[{{Abbott} {et~al.}(2016{\natexlab{c}}){Abbott}, {Abbott}, {Abbott},
  {Abernathy}, {Acernese}, {Ackley}, {Adams}, {Adams}, {Addesso}, {Adhikari},
  \& et~al.}]{PEpaper}
---. 2016{\natexlab{c}},
  \href{http://dx.doi.org/10.1103/PhysRevLett.116.241102}{\JournalTitle{Physical
  Review Letters}, 116, 241102}

\bibitem[{{Antonini} \& {Rasio}(2016)}]{Antonini16}
{Antonini}, F., \& {Rasio}, F.~A. 2016,
  \href{http://dx.doi.org/10.3847/0004-637X/831/2/187}{\JournalTitle{\apj},
  831, 187}

\bibitem[{{Belczynski} {et~al.}(2016){Belczynski}, {Holz}, {Bulik}, \&
  {O'Shaughnessy}}]{2016Natur.534..512B}
{Belczynski}, K., {Holz}, D.~E., {Bulik}, T., \& {O'Shaughnessy}, R. 2016,
  \href{http://dx.doi.org/10.1038/nature18322}{\JournalTitle{\nat}, 534, 512}

\bibitem[{{Berti} \& {Volonteri}(2008)}]{Berti}
{Berti}, E., \& {Volonteri}, M. 2008,
  \href{http://dx.doi.org/10.1086/590379}{\JournalTitle{\apj}, 684, 822}

\bibitem[{{Bird} {et~al.}(2016){Bird}, {Cholis}, {Mu{\~n}oz},
  {Ali-Ha{\"i}moud}, {Kamionkowski}, {Kovetz}, {Raccanelli}, \&
  {Riess}}]{2016PhRvL.116t1301B}
{Bird}, S., {Cholis}, I., {Mu{\~n}oz}, J.~B., {et~al.} 2016,
  \href{http://dx.doi.org/10.1103/PhysRevLett.116.201301}{\JournalTitle{Physical
  Review Letters}, 116, 201301}

\bibitem[{{Buonanno} {et~al.}(2008){Buonanno}, {Kidder}, \& {Lehner}}]{BKL07}
{Buonanno}, A., {Kidder}, L.~E., \& {Lehner}, L. 2008,
  \href{http://dx.doi.org/10.1103/PhysRevD.77.026004}{\JournalTitle{\prd}, 77,
  026004}

\bibitem[{{Clesse} \& {Garc{\'{\i}}a-Bellido}(2017)}]{CG16}
{Clesse}, S., \& {Garc{\'{\i}}a-Bellido}, J. 2017,
  \href{http://dx.doi.org/10.1016/j.dark.2016.10.002}{\JournalTitle{Physics of
  the Dark Universe}, 15, 142}

\bibitem[{{de Mink} \& {Mandel}(2016)}]{2016MNRAS.460.3545D}
{de Mink}, S.~E., \& {Mandel}, I. 2016,
  \href{http://dx.doi.org/10.1093/mnras/stw1219}{\JournalTitle{\mnras}, 460,
  3545}

\bibitem[{{Gerosa} \& {Berti}(2017)}]{Gerosa}
{Gerosa}, D., \& {Berti}, E. 2017, \JournalTitle{ArXiv e-prints},
  \href{http://arxiv.org/abs/1703.06223}{{\sffamily arXiv:1703.06223 [gr-qc]}}

\bibitem[{{Healy} {et~al.}(2014){Healy}, {Lousto}, \& {Zlochower}}]{HLZ14}
{Healy}, J., {Lousto}, C.~O., \& {Zlochower}, Y. 2014,
  \href{http://dx.doi.org/10.1103/PhysRevD.90.104004}{\JournalTitle{\prd}, 90,
  104004}

\bibitem[{{Hofmann} {et~al.}(2016){Hofmann}, {Barausse}, \& {Rezzolla}}]{HBR16}
{Hofmann}, F., {Barausse}, E., \& {Rezzolla}, L. 2016,
  \href{http://dx.doi.org/10.3847/2041-8205/825/2/L19}{\JournalTitle{\apjl},
  825, L19}

\bibitem[{{Hogg} {et~al.}(2010){Hogg}, {Myers}, \& {Bovy}}]{Hogg2010}
{Hogg}, D.~W., {Myers}, A.~D., \& {Bovy}, J. 2010,
  \href{http://dx.doi.org/10.1088/0004-637X/725/2/2166}{\JournalTitle{\apj},
  725, 2166}

\bibitem[{{Hughes} \& {Blandford}(2003)}]{BH03}
{Hughes}, S.~A., \& {Blandford}, R.~D. 2003,
  \href{http://dx.doi.org/10.1086/375495}{\JournalTitle{\apjl}, 585, L101}

\bibitem[{{Inayoshi} {et~al.}(2017){Inayoshi}, {Hirai}, {Kinugawa}, \&
  {Hotokezaka}}]{2017arXiv170104823I}
{Inayoshi}, K., {Hirai}, R., {Kinugawa}, T., \& {Hotokezaka}, K. 2017,
  \JournalTitle{ArXiv e-prints},
  \href{http://arxiv.org/abs/1701.04823}{{\sffamily arXiv:1701.04823
  [astro-ph.HE]}}

\bibitem[{Jim{\'e}nez-Forteza {et~al.}(2017)Jim{\'e}nez-Forteza, Keitel, Husa,
  Hannam, Khan, \& Pürrer}]{Keitel}
Jim{\'e}nez-Forteza, X., Keitel, D., Husa, S., {et~al.} 2017,
  \href{http://dx.doi.org/10.1103/PhysRevD.95.064024}{\JournalTitle{Phys.
  Rev.}, D95, 064024}

\bibitem[{{Kesden}(2008)}]{Kesden08}
{Kesden}, M. 2008,
  \href{http://dx.doi.org/10.1103/PhysRevD.78.084030}{\JournalTitle{\prd}, 78,
  084030}

\bibitem[{{Kesden} {et~al.}(2010){Kesden}, {Sperhake}, \& {Berti}}]{Kesden10}
{Kesden}, M., {Sperhake}, U., \& {Berti}, E. 2010,
  \href{http://dx.doi.org/10.1103/PhysRevD.81.084054}{\JournalTitle{\prd}, 81,
  084054}

\bibitem[{{Kovetz} {et~al.}(2016){Kovetz}, {Cholis}, {Breysse}, \&
  {Kamionkowski}}]{2016arXiv161101157K}
{Kovetz}, E.~D., {Cholis}, I., {Breysse}, P.~C., \& {Kamionkowski}, M. 2016,
  \JournalTitle{ArXiv e-prints},
  \href{http://arxiv.org/abs/1611.01157}{{\sffamily arXiv:1611.01157}}

\bibitem[{{Lousto} {et~al.}(2010){Lousto}, {Nakano}, {Zlochower}, \&
  {Campanelli}}]{LoustoStatistics}
{Lousto}, C.~O., {Nakano}, H., {Zlochower}, Y., \& {Campanelli}, M. 2010,
  \href{http://dx.doi.org/10.1103/PhysRevD.81.084023}{\JournalTitle{\prd}, 81,
  084023}

\bibitem[{{Mandel} {et~al.}(2011){Mandel}, {Narayan}, \&
  {Kirshner}}]{Mandel2011}
{Mandel}, K.~S., {Narayan}, G., \& {Kirshner}, R.~P. 2011,
  \href{http://dx.doi.org/10.1088/0004-637X/731/2/120}{\JournalTitle{\apj},
  731, 120}

\bibitem[{{Mapelli}(2016)}]{Mapelli16}
{Mapelli}, M. 2016,
  \href{http://dx.doi.org/10.1093/mnras/stw869}{\JournalTitle{\mnras}, 459,
  3432}

\bibitem[{{McKernan} {et~al.}(2017){McKernan}, {Ford}, {Bellovary}, {Leigh},
  {Haiman}, {Kocsis}, {Lyra}, {MacLow}, {Metzger}, {O'Dowd}, {Endlich}, \&
  {Rosen}}]{McKernan}
{McKernan}, B., {Ford}, K.~E.~S., {Bellovary}, J., {et~al.} 2017,
  \JournalTitle{ArXiv e-prints},
  \href{http://arxiv.org/abs/1702.07818}{{\sffamily arXiv:1702.07818
  [astro-ph.HE]}}

\bibitem[{{Merritt} {et~al.}(2004){Merritt}, {Milosavljevi{\'c}}, {Favata},
  {Hughes}, \& {Holz}}]{Merritt}
{Merritt}, D., {Milosavljevi{\'c}}, M., {Favata}, M., {Hughes}, S.~A., \&
  {Holz}, D.~E. 2004,
  \href{http://dx.doi.org/10.1086/421551}{\JournalTitle{\apjl}, 607, L9}

\bibitem[{{Miller} \& {Miller}(2015)}]{MillerMiller}
{Miller}, M.~C., \& {Miller}, J.~M. 2015,
  \href{http://dx.doi.org/10.1016/j.physrep.2014.09.003}{\JournalTitle{\physrep},
  548, 1}

\bibitem[{{O'Leary} {et~al.}(2016){O'Leary}, {Meiron}, \&
  {Kocsis}}]{2016ApJ...824L..12O}
{O'Leary}, R.~M., {Meiron}, Y., \& {Kocsis}, B. 2016,
  \href{http://dx.doi.org/10.3847/2041-8205/824/1/L12}{\JournalTitle{\apjl},
  824, L12}

\bibitem[{{P{\"u}rrer} {et~al.}(2016){P{\"u}rrer}, {Hannam}, \&
  {Ohme}}]{Purrer}
{P{\"u}rrer}, M., {Hannam}, M., \& {Ohme}, F. 2016,
  \href{http://dx.doi.org/10.1103/PhysRevD.93.084042}{\JournalTitle{\prd}, 93,
  084042}

\bibitem[{{Rodriguez} {et~al.}(2016{\natexlab{a}}){Rodriguez}, {Chatterjee}, \&
  {Rasio}}]{2016PhRvD..93h4029R}
{Rodriguez}, C.~L., {Chatterjee}, S., \& {Rasio}, F.~A. 2016{\natexlab{a}},
  \href{http://dx.doi.org/10.1103/PhysRevD.93.084029}{\JournalTitle{\prd}, 93,
  084029}

\bibitem[{{Rodriguez} {et~al.}(2016{\natexlab{b}}){Rodriguez}, {Haster},
  {Chatterjee}, {Kalogera}, \& {Rasio}}]{2016ApJ...824L...8R}
{Rodriguez}, C.~L., {Haster}, C.-J., {Chatterjee}, S., {Kalogera}, V., \&
  {Rasio}, F.~A. 2016{\natexlab{b}},
  \href{http://dx.doi.org/10.3847/2041-8205/824/1/L8}{\JournalTitle{\apjl},
  824, L8}

\bibitem[{{Sigurdsson} \& {Hernquist}(1993)}]{1993Natur.364..423S}
{Sigurdsson}, S., \& {Hernquist}, L. 1993,
  \href{http://dx.doi.org/10.1038/364423a0}{\JournalTitle{\nat}, 364, 423}

\bibitem[{{Stevenson} {et~al.}(2017){Stevenson}, {Berry}, \&
  {Mandel}}]{Stevenson17}
{Stevenson}, S., {Berry}, C.~P.~L., \& {Mandel}, I. 2017, \JournalTitle{ArXiv
  e-prints}, \href{http://arxiv.org/abs/1703.06873}{{\sffamily arXiv:1703.06873
  [astro-ph.HE]}}

\bibitem[{{Tichy} \& {Marronetti}(2008)}]{Tichy08}
{Tichy}, W., \& {Marronetti}, P. 2008,
  \href{http://dx.doi.org/10.1103/PhysRevD.78.081501}{\JournalTitle{\prd}, 78,
  081501}

\bibitem[{{Vitale} {et~al.}(2017){Vitale}, {Lynch}, {Sturani}, \&
  {Graff}}]{Vitale15}
{Vitale}, S., {Lynch}, R., {Sturani}, R., \& {Graff}, P. 2017,
  \href{http://dx.doi.org/10.1088/1361-6382/aa552e}{\JournalTitle{Classical and
  Quantum Gravity}, 34, 03LT01}

\bibitem[{{Vitale} {et~al.}(2014){Vitale}, {Lynch}, {Veitch}, {Raymond}, \&
  {Sturani}}]{2014PhRvL.112y1101V}
{Vitale}, S., {Lynch}, R., {Veitch}, J., {Raymond}, V., \& {Sturani}, R. 2014,
  \href{http://dx.doi.org/10.1103/PhysRevLett.112.251101}{\JournalTitle{Physical
  Review Letters}, 112, 251101}

\end{thebibliography}

\end{document}